\pdfoutput=1
\documentclass[runningheads]{llncs}
\usepackage{graphicx}
\usepackage{cite}  %
\usepackage{letltxmacro}  %
\usepackage{amssymb}%
\usepackage{amsmath}%
\usepackage{footnote}
\usepackage{verbatim}
\usepackage{color}%

\usepackage{eso-pic}
\newcommand\AtPageUpperMyright[1]{\AtPageUpperLeft{%
 \put(\LenToUnit{0.3\paperwidth},\LenToUnit{-1.5cm}){%
     \parbox{0.8\textwidth}{\raggedleft\fontsize{9}{11}\selectfont #1}}%
 }}%
\newcommand{\conf}[1]{%
\AddToShipoutPictureBG*{%
\AtPageUpperMyright{#1}
}
}

\begin{document}
\title{An automatic COVID-19 CT segmentation network using spatial and channel attention mechanism}


\titlerunning{COVID-19 CT segmentation network}
\author{Tongxue Zhou\inst{1,2} \and
St\'ephane Canu\inst{3}
\and Su Ruan\inst{1}}
\authorrunning{Zhou, Canu, Vera, Ruan.}
\institute{Universit\'e de Rouen Normandie, LITIS - QuantIF, Rouen 76183, France
\and
INSA Rouen, LITIS - Apprentissage, Rouen 76800, France \and 
Normandie Univ, INSA Rouen, UNIROUEN, UNIHAVRE, LITIS, France}

\maketitle  
\conf{\textbf{Published in International Journal of Imaging Systems and Technology (2020). https://doi.org/10.1002/ima.22527}}

\begin{abstract}
The coronavirus disease (COVID-19) pandemic has led to a devastating effect on the global public health. Computed Tomography (CT) is an effective tool in the screening of COVID-19. It is of great importance to rapidly and accurately segment COVID-19 from CT to help diagnostic and patient monitoring. In this paper, we propose a U-Net based segmentation network using attention mechanism. As not all the features extracted from the encoders are useful for segmentation, we propose to incorporate an attention mechanism including a spatial and a channel attention, to a U-Net architecture to re-weight the feature representation spatially and channel-wise to capture rich contextual relationships for better feature representation. In addition, the focal tversky loss is introduced to deal with small lesion segmentation. The experiment results, evaluated on a COVID-19 CT segmentation dataset where 473 CT slices are available, demonstrate the proposed method can achieve an accurate and rapid segmentation on COVID-19 segmentation. The method takes only 0.29 second to segment a single CT slice. The obtained Dice Score, Sensitivity and Specificity are 83.1\%, 86.7\% and 99.3\%, respectively.
\end{abstract}

\section{Introduction}

In December 2019, a novel coronavirus, now designated as COVID-19 by the World Health Organization (WHO), was identified as the cause of an outbreak of acute respiratory illness \cite{wu2020nowcasting, world2020director}. The pandemic of COVID-19 is spreading all over the world and causes a devastating effect on the global public health. As a form of pneumonia, the infection causes inflammation in alveoli, which fills with fluid or pus, making the patient difficult to breathe \cite{shi2020large}. Similar to other coronaviral pneumonia such as Severe Acute Respiratory Syndrome (SARS) and Middle East Respiratory Syndrome (MERS), COVID-19 can also lead to acute respiratory distress syndrome (ARDS) \cite{huang2020clinical, li2020coronavirus}. In addition, the number of people infected by the virus is increasing rapidly. Up to April 19, 2020, 2,241,359 cases of COVID-19 have been reported in over 200 countries and territories, resulting in approximately 152,551 deaths\footnote{https://www.who.int/emergencies/diseases/novel-coronavirus-2019/situation-reports/}, while there is no efficient treatment at present. 

Due to the fast progression and infectious ability of the disease, it's urgent to develop some tools to accurate diagnose and evaluate the disease. Although the real-time polymerase chain reaction (RT-PCR) assay of the sputum is considered as the gold standard for diagnosis, while it is time-consuming and has been reported to suffer from high false negative rates \cite{shan+2020lung, liang2020handbook}. In clinical practice, Chest Computed tomography (CT), as a non-invasive imaging approach, can detect certain characteristic manifestations in the lung associated with COVID-19, for example, ground-glass opacities and consolidation are the most relative imaging features in pneumonia associated with SARS-CoV-2 infection. Therefore, Chest CT is considered as a low-cost, accurate and efficient method diagnostic tool for early screening and diagnosis of COVID-19. It can be evaluated how severely the lungs are affected, and how the patient’s disease is evolving, which is helpful in making treatment decisions \cite{li2020artificial,pan2020time,ng2020imaging,lei2020ct, ye2020chest}. 

A number of artificial intelligence (AI) systems based on deep learning have been proposed and results have been shown to be quite promising in medical image analysis \cite{hu2020effective,piantadosi2020multi, bui2019incorporated, savelli2020multi}.
Compared to the traditional imaging workflow  heavily relies on the human labors, AI enables more safe, accurate and efficient imaging solutions. Recent AI-empowered applications in COVID-19 mainly include the dedicated imaging platform, the lung and infection region segmentation, the clinical assessment and diagnosis, as well as the pioneering basic and clinical research \cite{shi2020review}. Segmentation is an essential step in AI-based COVID-19 image processing and analysis for make a prediction of disease evolution. It delineates the regions of interest (ROIs), e.g., lung, lobes, bronchopulmonary segments, and infected regions or lesions, in the chest X-ray or CT images for further assessment and quantification \cite{shi2020review}. There are a number of researches related to COVID-19. For example, Zheng et al. \cite{zheng2020deep} proposed a weakly-supervised deep learning-based software system using 3D CT volumes to detect COVID-19. Goze et al. \cite{gozes2020rapid} presented a system that utilises 2D slice analysis and 3D volume analysis to achieve the detection of COVID-19. Jin et al. \cite{jin2020ai} proposed an AI system for fast COVID-19 diagnosis, where a segmentation model is first used to obtain the lung lesion regions, and then the classification model is used to determine whether it is COVID-19-like for each lesion region. Li et al. \cite{li2020artificial} developed a COVID-19 detection neural network (COVNet) to extract visual features from volumetric chest CT exams for distinguishing COVID-19 from Community Acquired Pneumonia (CAP). Chen et al. \cite{chen2020deep} proposed to use UNet++\cite{zhou2018unet++} to extract valid areas and detect suspicious lesions in CT images.

U-net \cite{ronneberger2015u} is the most widely used encoder-decoder network architecture for medical image segmentation, since the encoder captures the low-level and high-level features, and the decoder combines the semantic features to construct the final result. However, not all features extracted from the encoder are useful for segmentation. Therefore, it is necessary to find an effective way to fuse features, we focus on the extraction of the most informative features for segmentation. Hu et al. \cite{hu2018squeeze} introduced the Squeeze and Excitation (SE) block to improve the representational power of a network by modelling the interdependencies between the channels of its convolutional features. Roy et al. \cite{roy2018concurrent} introduced to use both spatial and channel SE blocks (scSE), which concurrently recalibrates the feature representations spatially and channel-wise, and then combine them to obtain the final feature representation. Inspired by this work, we incorporate an attention mechanism including both spatial attention and channel one to our segmentation network to extract more informative feature representation to enhance the network performance.

In this paper, we propose a deep learning based segmentation with the attention mechanism. A preliminary conference version appeared at ISBI 2020 \cite{zhou2020a}, which focused on the multi-model fusion issue. This journal version is a substantial extension, including (1) An automatic COVID-19 CT segmentation network. (2) A focal tversky loss function (different from the paper of ISBI) which is introduced to help to segment the small COVID-19 regions. (3) An attention mechanism including a spatial and a channel attention is introduced to capture rich contextual relationships for better feature representations.
The paper is organized as follows: Section \ref{sec2} offers an overview of this work and details our model, Section \ref{sec3} describes experimental setup, Section \ref{sec4} presents the experimental results, Section \ref{sec5} discusses the proposed method and concludes this work.

\section{Method}
\label{sec2}
\subsection{The proposed network architecture}
\begin{figure*}[htb]
\centering
\includegraphics[width=12cm]{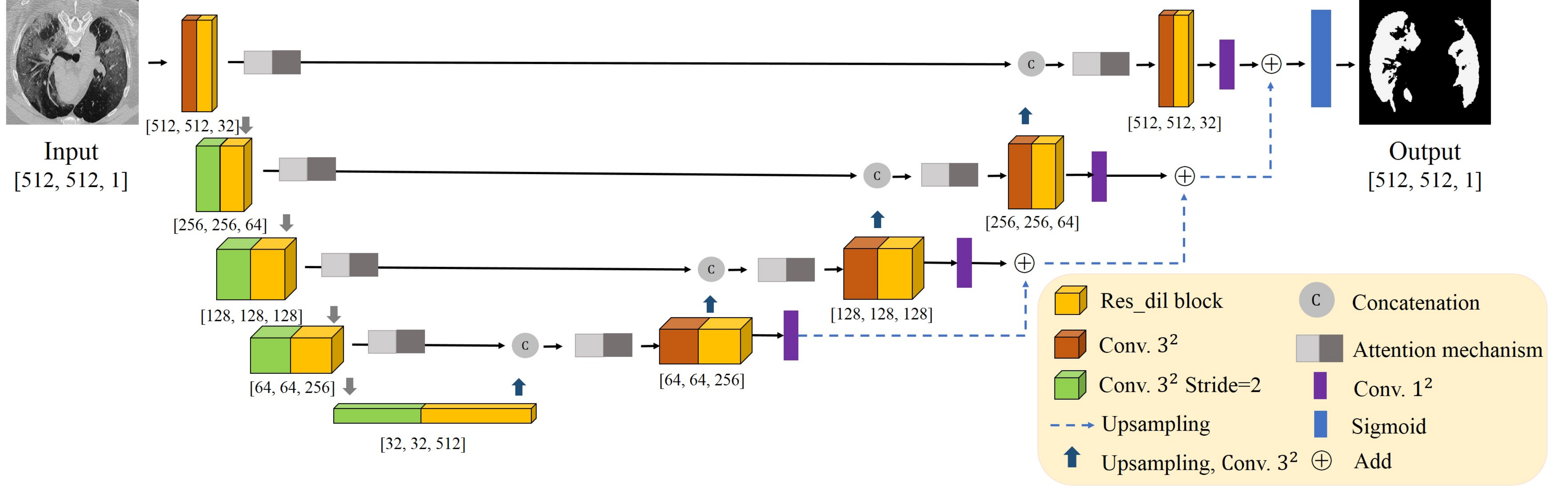}
\caption{The architecture of the proposed network. The network takes a CT slice as input and directly outputs the COVID-19 region.}
\label{Fig.1}
\end{figure*}

Our network is mainly based on the U-Net architecture \cite{ronneberger2015u}, in which we integrate an attention mechanism, res\_dil block and deep supervision. The encoder of the U-Net is used to obtain the feature representations. The feature representation at each layer are input into an attention mechanism, where they will be re-weighted along channel-wise and space-wise, and the most informative representations can be obtained, and finally they are projected by decoder to the label space to obtain the segmentation result. In the following, we will describe the main components of our model: encoder, decoder, and res\_dil block, deep supervision and attention mechanism. The network architecture scheme is described in Fig. \ref{Fig.1}.

\subsection{Encoder and decoder}
The encoder is used to obtain the feature representations. It includes a convolutional block, a res\_dil block followed by skip connection. In order to maintain the spatial information, we use a convolution with stride = 2 to replace pooling operation. It's likely to require different receptive field when segmenting different regions in an image. All convolutions are $3\times3$ and the number of filter is increased from 32 to 512. Each decoder level begins with upsampling layer followed by a convolution to reduce the number of features by a factor of 2. Then the upsampled features are combined with the features from the corresponding level of the encoder part using concatenation. After the concatenation, we use the res\_dil block to increase the receptive field. In addition, we employ deep supervision \cite{isensee2017brain} for the segmentation decoder by integrating segmentation layers from different levels to form the final network output, shown in Fig. \ref{Fig.2}.

\subsection{Res\_dil block and deep supervision}
\begin{figure}[htb]
\centering
\includegraphics[width=10cm]{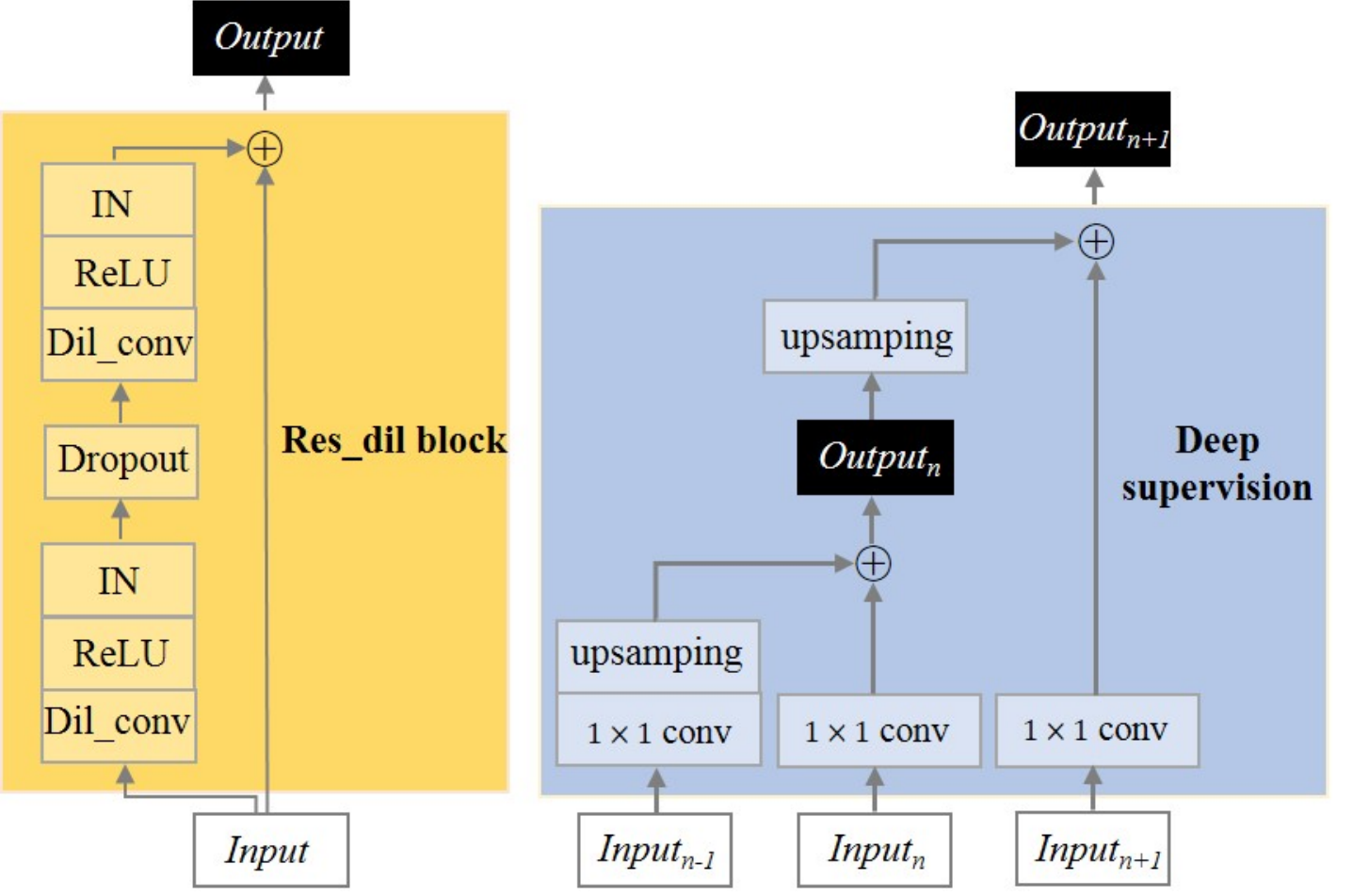}
\caption{The architecture of our proposed Res\_dil block (left) and Deep supervision (right). IN refers instance normalization, Dil\_conv the dilated convolution (rate = 2, 4, respectively). We refer to the vertical depth as level, with higher levels being higher spatial resolution. In the deep supervision part, $Input_n$ refers the output of res\_dil block of the $n_{th}$ level in the decoder, $Output_n$ refers the segmentation result of the $n_{th}$ level in the decoder.}
\label{Fig.2}
\end{figure}

It's likely to require different receptive field when segmenting different regions in an image. Since standard U-Net can not get enough semantic features due to the limited receptive field, inspired by dilated convolution \cite{yu2015multi}, we proposed to use residual block with dilated convolutions on both encoder part and decoder part to obtain features at multiple scales, the architecture of res\_dil is shown in Fig. \ref{Fig.2}. The res\_dil block can obtain more extensive local information to help retain information and fill details during training process.

To demonstrate that the proposed res\_dil can enlarge the receptive field mathematically, we let $F:Z^2 \rightarrow R$ be a  discrete function, $\Omega_r=[-r,r]^2\in Z^2$ and let $k:\Omega_r \rightarrow R$ be a discrete filter size $(2r+1)^2$. The discrete convolution operator $\star$ can be described as follows:

\begin{equation}
    (F\star k)=\sum_{m=-r}^{r}\sum_{n=-r}^r F(x-m,y-n)k(m,n)
\end{equation}

Let $l$ be a dilation factor and the $l$-dilated convolution operation $\star_l$ can be defined as:

\begin{equation}
    (F\star_lk)=\sum_{m=-r}^{r}\sum_{n=-r}^r F(x-lm,y-ln)k(m,n)
\end{equation}

We assume $F_0$, $F_1$,..., $F_{n-1} :$ $Z^2\rightarrow R$ are a discrete functions, and $k_0$, $k_1$,..., $k_{n-2} :$ $Z^2\rightarrow R$ are discrete $3\times 3$ filters. In addition, we apply the filters with exponentially increasing dilation factors, such as $2^0$,  $2^1$,... $2^{n-2}$. Then, the discrete function $F_{i+1}$ can be described as:

\begin{equation}
    F_{i+1}=F_i\star_{2^i}k_i, i=0,1,...,n-2
\end{equation}

\begin{figure}[htb]
\centering
\includegraphics[width=12cm]{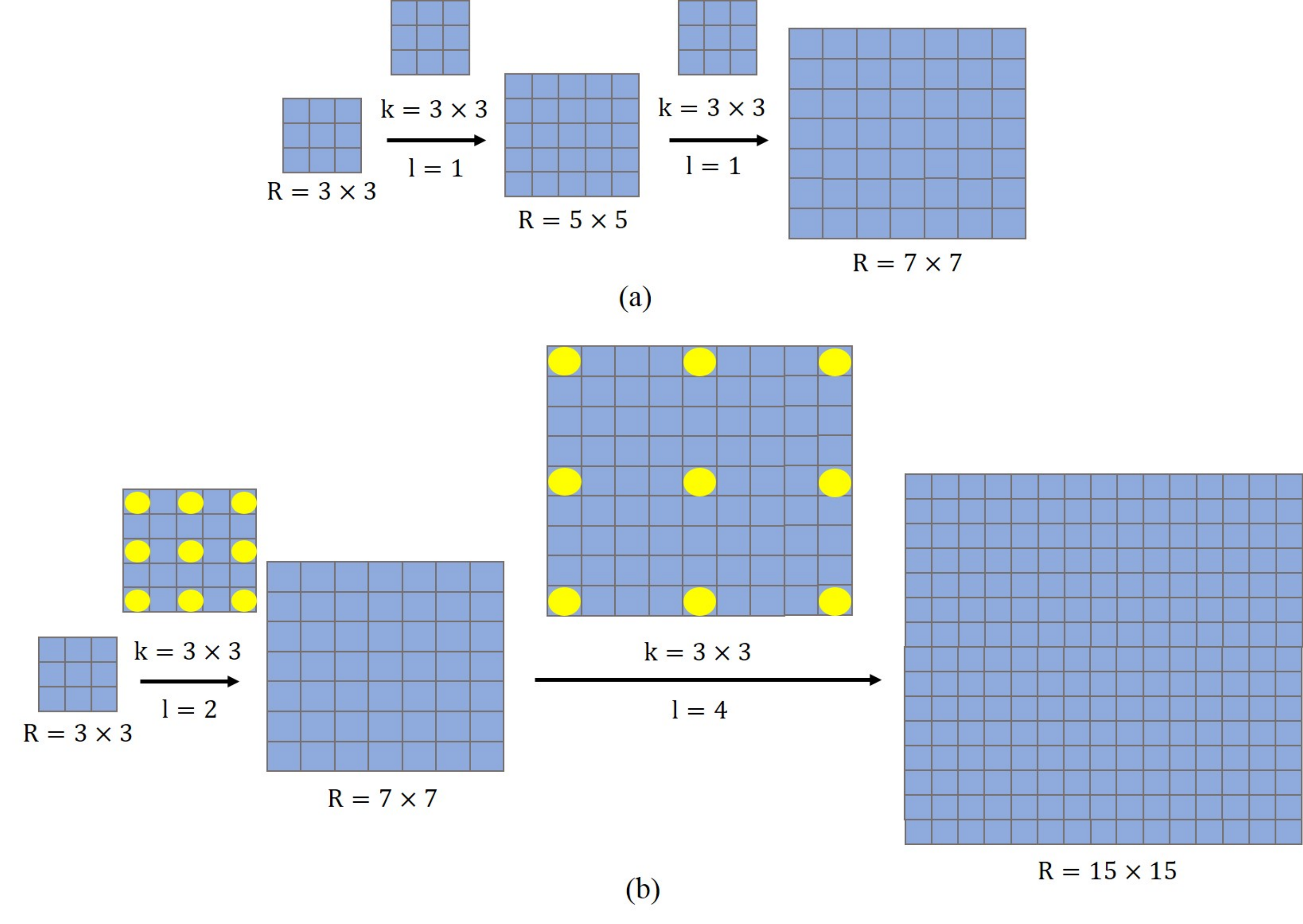}
\caption{The illustration of receptive field, $R$ denotes the receptive field, $k$ denotes the convolution kernel size, and $l$ denotes the dilated factor. (a) a convolution network which consists of two $k=3\times3$ and $l=1,1$ convolutional layers. (b) a convolution network which consists of two $k=3\times3$ and $l=2,4$ dilated convolutional layers.}
\label{Fig.3}
\end{figure}

According to the definition of receptive field, the receptive
field size of each element in $F_{i+1}$ is $(2^{i+2}-1)\times(2^{i+2}-1)$, which is a square of exponentially increasing size. So we can obtain a $15\times15$ receptive field by applying our proposed res\_dil block with the dilation factor 2 and 4, respectively, while the classical convolution can only obtain $7\times7$ receptive field, see Fig. \ref{Fig.3}.

\subsection{Attention mechanism}

\begin{figure}[htb]
\centering
\includegraphics[width=12cm]{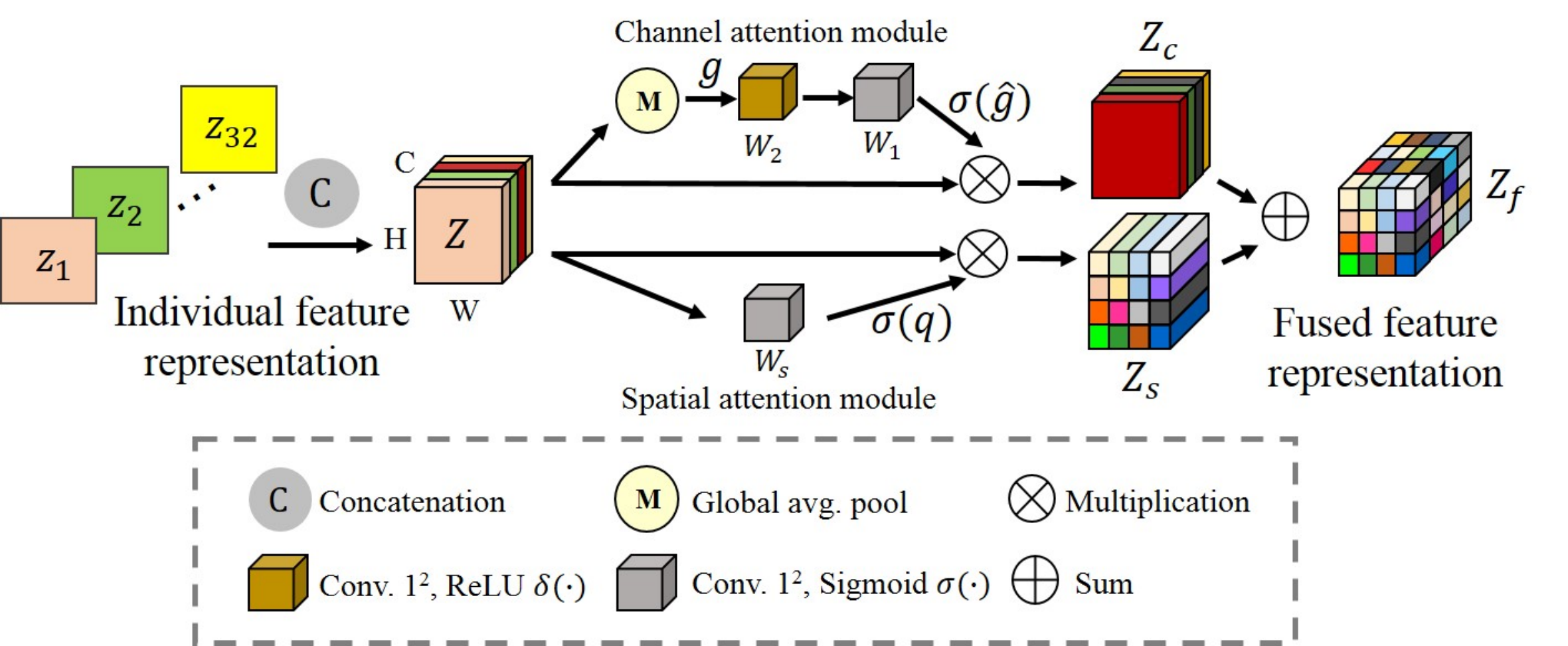}
\caption{The architecture of attention mechanism. The individual feature representations ($z_1$, $z_2$, ..., $z_{32}$) are first concatenated as $Z$, and then they are recalibrated spatially and channel-wise to achieve the $Z_s$ and $Z_c$, final they are added to obtain the rich fused feature representation $Z_f$.}
\label{Fig.4}
\end{figure}

In U-net shaped network, not all the features obtained by the encoder are effective for segmentation. In addition, not only the different channels (filters) have various contributions but also different spatial location in each channel can give different weights on feature representation for segmentation. To this end, we introduced a "scSE based " attention mechanism in both encoder and decoder to take into account the most informative feature representations along channel-wise and spatial-wise for segmentation, the architecture is described in Fig. \ref{Fig.4}. 

The individual feature representations from each channel are first concatenated as the input representation $Z= [z_1, z_2, ..., z_n]$, $Z_k\in R^{H\times W}$, $n$ is the number of channel in each layer. To simplify the description, we take $n=32$. 

In the channel attention module, a global average pooling is first performed to produce a tensor $g\in R^{1\times1\times 32}$, which represents the global spatial information of the representation, with its $k^{th}$ element
\begin{equation}
    g_k= \frac{1}{H\times W}\sum_i^H\sum_j^W Z_k(i,j)
\end{equation}

Then two fully-connected layers are applied to encode the channel-wise dependencies, $\hat{g}= W_1(\delta(W_2 g))$, with $W_1\in R^{32\times16}$, $W_2\in R^{16\times32}$, being weights of two fully-connected layers and the ReLU operator $\delta(\cdot)$. $\hat{g}$ is then passed through the sigmoid layer to obtain the channel-wise weights, which will be applied to the input representation $Z$ through multiplication to achieve the channel-wise representation $Z_c$, the $\sigma(\hat{g_k})$ indicates the importance of the $i$ channel of the representation: \begin{equation}
    Z_c= [\sigma (\hat{g_1})z_1, \sigma(\hat{g_2})z_2, ..., \sigma(\hat{g_{32}})z_{32},]
\end{equation}

In the spatial attention module, the representation can be considered as $Z= [z^{1,1}, z^{1,2}, ... , z^{i,j},..., z^{H,W}]$, $Z^{i,j}\in R^{1\times1\times32}$, $i\in{1, 2,..., H}$, $j\in{1, 2,...,W}$, and then a convolution operation $q=W_s\star Z$, $q\in R^{H\times W}$  with weight $W_s\in R^{1\times1\times32\times1}$, is used to squeeze the spatial domain, and to produce a projection tensor, which represents the linearly combined representation for all channels for a spatial location. The tensor is finally passed through a sigmoid layer to obtain the space-wise weights and to achieve the spatial-wise representation $Z_s$, the $\sigma(q_{i,j})$ that indicates the importance of the spatial information $(i, j)$ of the representation:
\begin{equation}
    Z_s=[\sigma(q_{1,1})z^{1,1},  ... , \sigma(q_{i,j})z^{i,j}, ... , \sigma(q_{H,W})z^{H,W}]
\end{equation}

The fused feature representation is obtained by adding the channel-wise representation and space-wise representation:
\begin{equation}
    Z_f=Z_c+Z_s 
\end{equation}

The attention mechanism can be directly adapted to any feature representation problem, and it encourages the network to capture rich contextual relationships for better feature representations.

\subsection{Loss function}
In the medical community, the Dice Score Coefficient (DSC), defined in (5), is the most widespread metric to measure the overlap ratio of the segmented region and the ground truth, and it is widely used to evaluate segmentation performance. Dice Loss (DL) in (6) is defined as a minimization of the overlap between the prediction and ground truth.

\begin{equation}
    \ DSC_c=\frac{\sum_{j=1}^N p_{ic} g_{ic}+\epsilon}{\sum_{j=1}^N p_{ic} + g_{ic}+\epsilon}
\end{equation}

\begin{equation}
    \ DL_c=\sum_c(1-DSC_c)
\end{equation}

\noindent where $N$ is the number of pixels in the image, $c$ is the set of the classes, $p_{ic}$ is the probability that pixel $i$ is of the tumor class $c$ and $p_{i\overline{c}}$ is the probability that pixel $i$ is of the non-tumor class $\overline{c}$. The same is true for $g_{ic}$ and $g_{i\overline{c}}$, and $\epsilon$ is a small constant to avoid dividing by 0.

One of the limitation of Dice Loss is that it penalizes false positive (FP) and false negative (FN) equally, which results in segmentation maps with high precision but low recall. This is particularly true for highly imbalanced dataset and small regions of interests (ROI) such as COVID-19 lesions. Experimental results show that FN needs to be weighted higher than FP to improve recall rate. Tversky similarity index \cite{tversky1977features} is a generalization of the DSC which allows for flexibility in balancing FP and FN:

\begin{equation}
    \ TI_c=\frac{\sum_{j=1}^N p_{ic} g_{ic}+\epsilon} {\sum_{j=1}^N p_{ic} g_{ic}+\alpha\sum_{j=1}^N p_{i\overline{c}} g_{ic}+\beta\sum_{j=1}^N p_{ic} g_{i\overline{c}}+\epsilon}
\end{equation}


Another issue with the DL is that it struggles to segment small ROIs as they do not contribute to the loss significantly. To address this, Abraham et al. \cite{abraham2019novel} proposed the Focal Tversky Loss function (FTL).

\begin{equation}
    \ {FTL_c}=\sum_c(1-TI_c)^{1/\gamma}
\end{equation}

\noindent where $\gamma$ varies in the range $[1, 3]$. In practice, if a pixel is misclassified with a high Tversky index, the FTL is unaffected. However, if the Tversky index is small and the pixel is misclassified, the FTL will decrease significantly. To this end, we used FTL to train the network to help segment the small COVID-19 regions.


\section{Experimental setup}
\label{sec3}
\subsection{Dataset and preprocessing}
\begin{figure}[htb]
\centering
\includegraphics[width=12cm]{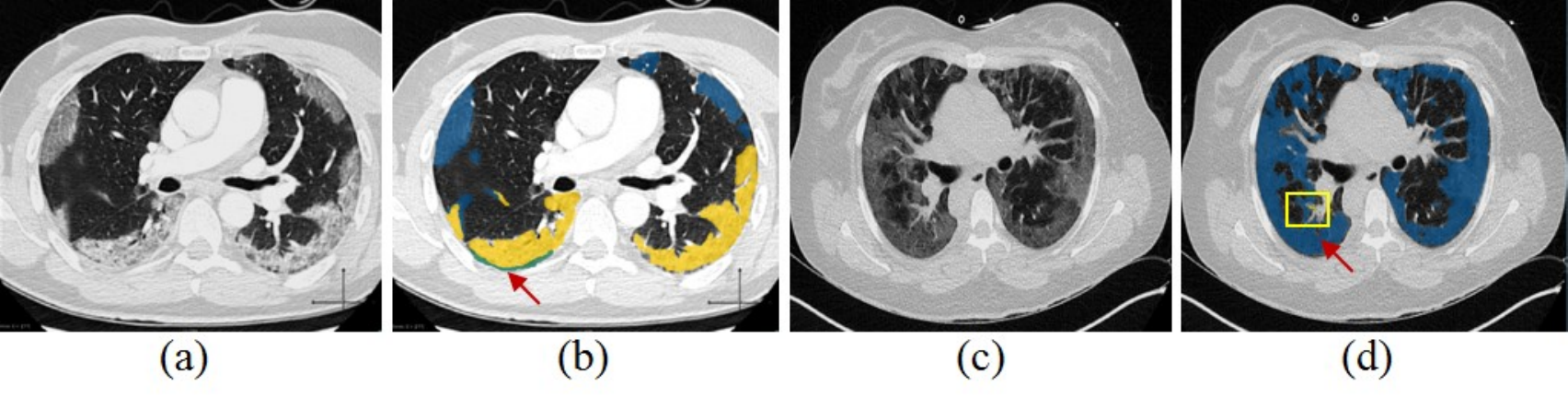}
\caption{Example images of the COVID-19 CT segmentation dataset. (a) and (c): CT image from Dataset-1 and Dataset-2, (b) and (d): The Ground truth of (a), (c), respectively, ground-glass is shown in blue, consolidation is shown in yellow and pleural effusion is shown in green.}
\label{Fig.5}
\end{figure}

The two datasets used in the experiments come from Italian Society of Medical and Interventional Radiology: COVID-19 CT segmentation dataset \footnote{http://medicalsegmentation.com/covid19/}. Dataset-1 includes 100 axial CT images from 60 patients with Covid-19. The images have been resized, greyscaled and compiled into a single NIFTI-file. The image size is $512\times512$ pixels. The images have been segmented by a radiologist using three labels: ground-glass, consolidation and pleural effusion. Dataset-2 includes 9 volumes, total 829 slices, where 373 slices have been evaluated and segmented by a radiologist as COVID-19 cases. We resize these images from $630\times630$ pixels to $512\times512$ pixels same as Dataset-1. And an intensity normalization is applied to both datasets. Since there are severe data imbalance in the dataset. For example, in Dataset-1, only 25 slices have pleural effusion, which is the smallest region among all the COVID-19 lesion regions (see the green region in Fig. \ref{Fig.4}). In Dataset-2, only 233 slices have consolidation, which takes up a small amount of pixels in the image (see the yellow region in Fig. \ref{Fig.4}). We take all the lesion labels as a COVID-19 lesion. Because of the small number of data in both two datasets, we combine the two datasets as our final training dataset. Here, we give some example images of the COVID-19 CT segmentation dataset in Fig. \ref{Fig.5}.

\subsection{Implementation details}
Our network is implemented in Keras with a single Nvidia GPU Quadro P5000 (16G). The network is trained by focal tversky loss and is optimized using the Adam optimizer, the initial learning rate = 5e-5 with a decreasing learning rate factor 0.5 with patience of 10 epochs. Early stopping is employed to avoid over-fitting if the validation loss is not improved over 50 epochs. We randomly split the dataset into 80\% training and 20\% testing.

\subsection{Evaluation metrics}

Segmentation accuracy determines the eventual success or failure of segmentation procedures. To measure the segmentation performance of the proposed methods, three evaluation metrics: Dice, Sensitivity and Specificity are used to obtain quantitative measurements of the segmentation accuracy.

\noindent 1) Dice: It is designed to evaluate the overlap rate of prediction results and ground truth. Dice ranges from 0 to 1, and the better predict result will have a larger Dice value.

\begin{equation}
 Dice = \frac {2TP}{2TP+FP+FN}
\end{equation}

\noindent 2) Sensitivity(also called the true positive rate, the recall): It measures the proportion of actual positives that are correctly identified:

\begin{equation}
Sensitivity=\frac{TP}{TP+FN}
\end{equation}

\noindent 3) Specificity(also called the true negative rate): It measures the proportion of actual negatives that are correctly identified:

\begin{equation}
Specificity =\frac{TN}{TN+FP}
\end{equation}

\noindent where $TP$ represents the number of true positive voxels, $TN$ represents the number of true negative voxels, $FP$ represents the number of false positive voxels, and $FN$ represents the number of false negative voxels. 




\section{Experiment results}
\label{sec4}
In this section, we conduct extensive comparative experiments including quantitative analysis and qualitative analysis to demonstrate the effectiveness of our proposed method.



\subsection{Quantitative analysis}
\label{4.1.1}
To assess the performance of our method, and to analyze the impact of the proposed components of our network, we did an ablation study with regard to the attention mechanism and Focal Tversky Loss function (FTL), we refer our proposed network without the attention mechanism to baseline, the results are shown in Table \ref{tab1}. We can observe the baseline trained with DL achieves dice score, sensitivity and specificity of 80.4\%, 75.7\%, 99.8\%, respectively. However, using the focal tversky loss can aide the network to focus more on the false negative voxels, which increases 0.87\% of dice score, 3.43\% of sensitivity for Baseline and 1.96\% of dice score, 13.04\% of sensitivity for our proposed network. We can also observe in Table \ref{tab1} that integrating the attention mechanism to the segmentation network can boost the performance, since we can see an increase of 1.37\% of dice score and 1.32\% of sensitivity for 'Baseline+DL' and 2.47\% of dice score and 10.73\% of sensitivity for 'Baseline+FTL'. The main reason is that the attention mechanism can help to emphasis on the most important feature representation for segmentation. In addition, the proposed network trained by FTL combines the benefits of attention mechanism with FTL to outperform all other methods with dice $=83.1\%$, sensitivity$=86.7\%$ and specificity$=99.3\%$.

\begin{table}[]
\centering
\caption{Comparison of different methods on COVID-19 CT segmentation dataset, Baseline denotes our proposed network without the attention mechanism, bold results show the best score.}
\vspace{0.5cm}
\label{tab1}
\resizebox{\textwidth}{!}{%
\begin{tabular}{@{}ccccc@{}}
\hline
Model & Parameters & Dice (\%) & Sensitivity (\%)& Specificity (\%) \\ 
\hline
Baseline + DL & {$\alpha=0.5$, $\beta=0.5$} &80.4&75.7 &\textbf{99.8}\\
Baseline + FTL &{$\alpha=0.7$, $\beta=0.3$, $\gamma=4/3$}& 81.1&78.3&99.7 \\
\hline
Ours + DL & {$\alpha=0.5$, $\beta=0.5$} & 81.5&76.7&99.7\\
Ours + FTL & {$\alpha=0.7$ $\beta=0.3$, $\gamma=4/3$} & \textbf{83.1} & \textbf{86.7} & 99.3 \\ 
\hline
\end{tabular}%
}
\end{table}

\subsection{Qualitative analysis}
\label{4.2}
\begin{figure}[htb]
\centering
\includegraphics[width=12cm]{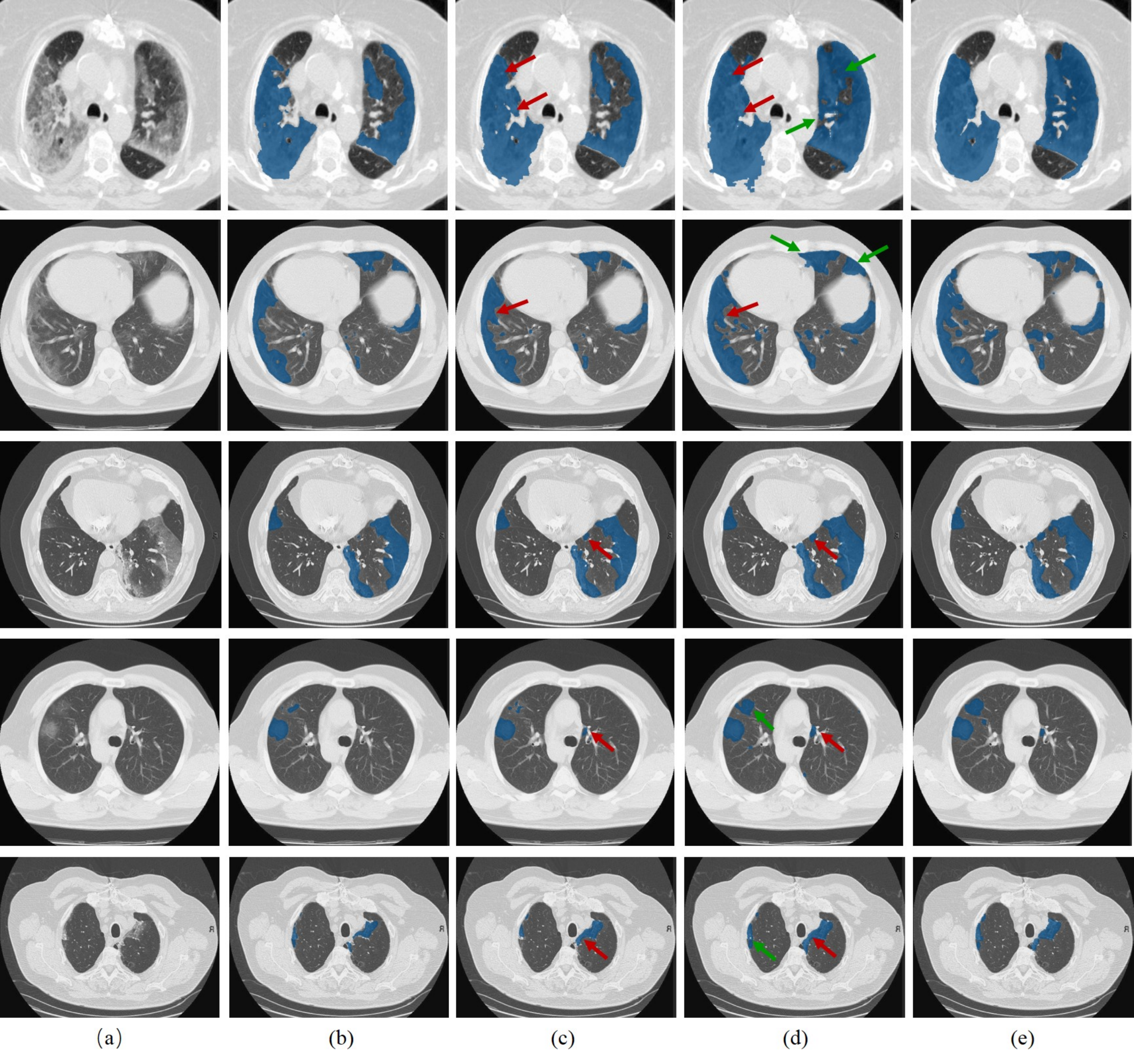}
\caption{Segmentation results of some examples on COVID-19 CT dataset. The first three examples are with many COVID-19 lesion regions, the last two examples are with few COVID-19 lesion regions. (a) CT image, (b) Baseline trained by dice loss, (c) Baseline trained by focal tversky loss, (d) Proposed network trained by focal tversky loss, (e) Ground truth, red arrow emphasizes the improvement of using focal tversky loss (from (b) to (c)), green arrow emphasizes the improvement of applying attention mechanism (from (c) to (d)).}
\label{Fig.6}
\end{figure}

In order to evaluate the effectiveness of our model, we randomly select several examples on COVID-19 CT segmentation dataset and visualize the results in Fig. \ref{Fig.6}. From Fig. \ref{Fig.6}, we can observe that the baseline trained by DL could give a rough segmentation result, while it fails to segment many small lesion regions. With the application of focal tversky loss, it can help to improve the segmentation result with a much better result. In addition, the attention mechanism can help to further refine the segmentation result. The proposed network trained by FTL can achieve the result closest to the ground truth. The obtained results have demonstrated that leveraging the attention mechanism and FTL can generally enhance the COVID-19 segmentation performance.

\section{Conclusion}
\label{sec5}
In this paper, we have presented a U-Net based segmentation network using attention mechanism. Since most current segmentation networks are trained with dice loss, which penalize the false negative voxels and false positive voxels equally, contributing a high specificity but low sensitivity. To this end, we applied the focal tversky loss to train the model to improve the small ROI segmentation performance. Moreover, we improve the baseline by incorporating an attention mechanism including a spatial attention and a channel attention in each layer to capture rich contextual relationships for better feature representations. We evaluated our proposed network on COVID-19 CT segmentation datasets, and the experiment results demonstrate the the superior performance of our method. However, the study is limited by the small dataset, in the future we would like to apply a larger training dataset to refine our model, and achieve more competitive results.
\vfill
\pagebreak
\bibliographystyle{splncs04}
\bibliography{strings}
\end{document}